\documentclass[11pt,twoside]{article}

\usepackage{asp2006}
\usepackage{epsf}
\usepackage{graphicx}
\usepackage{lscape}

\begin{document}
\title{How Compact are the Cores of AGN  ? Sub-Parsec Scale Imaging with VLBI at Millimeter Wavelength}   %%% Fill in title
\author{T.P. Krichbaum\altaffilmark{1}, S.S. Lee\altaffilmark{1}, A.P. Lobanov\altaffilmark{1}, A.P. Marscher\altaffilmark{2}, M.A. Gurwell\altaffilmark{3}}   %%% Fill in author names
%\altaffiltext{1}{Max-Planck-Institut f\"ur Radioastronomie, Bonn, Germany}
%\altaffiltext{2}{Institute for Astrophysical Research, Boston University, Boston, MA 02215, USA}
%\altaffiltext{3}{Harvard-Smithsonian Center for Astrophysics, 60 Garden Street, Cambridge, MA 02138, USA}
\affil{$^1$Max-Planck-Institut f\"ur Radioastronomie, Bonn, Germany}    %%% Fill in author affiliations
\affil{$^2$Institute for Astrophysical Research, Boston University, Boston, MA 02215, USA}    %%% Fill in author affiliations
\affil{$^3$Harvard-Smithsonian Center for Astrophysics, Cambridge, MA 02138, USA}    %%% Fill in author affiliations

\begin{abstract} %%% Abstract to run on from here.
We study the most central regions of AGN jets with an angular resolution of tens of micro-arcseconds using VLBI
at millimeter wavelengths (mm-VLBI).
We present and discuss a new 86\,GHz VLBI survey of compact radio sources. We show new high dynamic range images
of two nearby radio galaxies (3C\,120 and M\,87). In M\,87 the size of the compact VLBI core (the jet base) 
is $< 15$ Schwarzschild radii. Future mm-VLBI observations at 1\,mm and shorter wavelengths should lead to 
images of galactic and extragalactic radio sources with a spatial resolution down to a few
Schwarzschild radii of the central super massive black holes. To achieve this, the participation of
large and sensitive millimeter and sub-millimeter telescopes in VLBI is essential.

\end{abstract}

%%% MAIN BODY OF TEXT GOES HERE. CONSULT "INSTRUCTIONS FOR AUTHORS USING
%%% LATEX2E MARKUP", SECTIONS 2.3-2.6 FOR HELP WITH EQUATIONS, FIGURES,
%%% AND TABLES.

%\section{}   %%% Top level section head (remove "%" symbol)
%\subsection{}   %%% Second level section head (remove "%" symbol)
%\subsubsection{}   %%% Lowest level section head (remove "%" symbol)
%\section*{}    %%% Unnumbered top level section head (remove "%" symbol)
%\subsection*{}   %%% Unnumbered second level section head (remove "%" symbol)

\vspace*{-1.0cm}
\section{Introduction and Scientific Motivation}
Radio galaxies and quasars produce powerful relativistic plasma jets, accelerating particles
up to TeV energies. Jets are not unique to AGN, but are also observed in other objects like 
in X-ray binaries, micro-quasars, young stellar objects and probably also in Gamma-ray bursters.
In modern theory, the launching and acceleration of the relativistic jets is explained via MHD
processes, with magnetic fields either attached to the rotating accretion disk or directly linked to 
the ergosphere of the central and possibly rotating super-massive black hole \cite[e.g.][]{Narayan05}.

In order to better understand the process of jet formation, a study of the inner most
regions of compact radio sources is necessary.
VLBI at short millimeter wavelengths (mm-VLBI) provides, via its unrivaled angular resolution of
only a few ten micro-arcseconds (at $\lambda =1$\,mm), an almost unique tool to test and discriminate
between the different theoretical models. In addition to the high spatial resolution
(corresponding to $\leq 10$ Schwarzschild radii for nearby sources like Sgr\,A* or M\,87),
the high observing frequency allows to image emission regions, which are self-absorbed and
therefore not observable at longer wavelengths.

In the following we present some results from global 3\,mm-VLBI performed with the 
Global Millimeter VLBI Array (GMVA\footnote{see: http://www.mpifr-bonn.mpg.de/div/vlbi/globalmm/index.html}) at 86\,GHz.
VLBI observations at shorter wavelengths ($\lambda < 3$\,mm, $\nu > 100$\,GHz) are possible but
are still limited by the number of available mm-telescopes (typically $N \leq  3-4$), their sensitivity and the limited
observing bandwidth \citep[see e.g.][and references therein]{Krichbaum07}. 
To date, it is therefore not yet possible to produce a reliable map of a radio source at these shorter wavelengths.
The GMVA combines the sensitive and mm-capable telescopes in Europe 
%(100\,m Effelsberg --  MPIfR, Germany , 6x15\,m Plateau de Bure interferometer -- IRAM, France, 
%30\,m Pico Veleta -- IRAM, Spain, 20\,m Onsala -- OSO, Sweden and 14\,m 
%Mets\"ahovi -- MRO, Finland) with the VLBA (NRAO, USA). 
(100\,m Effelsberg, 6x15\,m Plateau de Bure, 30\,m Pico Veleta, 20\,m Onsala, \& 14\,m Mets\"ahovi) with the VLBA.
The GMVA is open to the scientific community and calls for observing 
proposals twice per year (proposal deadlines: Feb.\ $1^{st}$, Oct.\ $1^{st}$).
It enables imaging of compact radio sources at an angular resolution of $\ga 40 \mu$as. The transatlantic detection
sensitivity ($7\sigma$) ranges between $\sim (80-200)$\,mJy on the most sensitive baselines (for 512\,Mbps). 
This should be compared with the
$\sim (300-400)$\,mJy baseline sensitivity of the stand-alone VLBA. In a typical 12\,hr GMVA observing run, an image  
rms of $\sigma_{\rm map} \la (1-4)$\,mJy is reached.

\begin{table}[t]
{\small
\begin{center}
\begin{tabular}{lcclc} 
     Author                       &  Ant.       & Sensitivity   &   Number     &   Detections     \\ \hline
     Beasley et al. 1997          &    3        & $\sim$ 0.5    &   N=45       &   16\,\%   \\
     Londsdale et al. 1998        &    3-5      & $\sim$ 0.7    &   N=79       &   14\,\%   \\
     Rantakyro et al. 1998        &    6-9      & $\sim$ 0.5    &   N=67       &   24\,\%   \\
     Lobanov et al. 2000          &    3-5      & $\sim$ 0.4    &   N=28       &   93\,\%   \\
     Lee et al. 2008              &   12        & $\sim$ 0.2    &   N=127      &   95\,\%   \\
\end{tabular}
\end{center}
}
\vspace{-0.5cm}
\caption{
VLBI surveys at 86\,GHz: Col.1 gives the reference, col.2 the number of antennas,
col.3 the typical detection sensitivity ($7\sigma$) in [Jy], col.4 the number of observed sources and col.5 
the percentage of detected objects (detection rate).
}
\label{survey}
\vspace{-0.5cm}
\end{table}

\vspace*{-0.5cm}
\section{A New and Comprehensive 86 GHz VLBI Survey}

VLBI surveys at the highest possible frequency and resolution are important for
answering questions on the compactness of extragalactic radio sources,
systematic differences between source classes, and if and how the brightness temperature
varies along the jet and towards the region, where the jet is made. The latter
is closely related to observational tests of jet launching models.  Due to the limited sensitivity of early
3\,mm-VLBI and the relatively small number of telescopes,
previous surveys revealed only moderately low ($\leq 20-30$\,\%) detection rates and
small numbers of observable sources. Table \ref{survey} summarizes the existing 
3\,mm VLBI surveys in chronological order.

After more sensitive VLBI telescopes became available around year 2000,
a new and larger than previous VLBI snap-short survey was conducted. The observations
took place in 3 sessions (October 2001, April \& October 2002), with the participation
of the VLBA (8 antennas), Haystack, Pico Veleta, Plateau de Bure, 
Effelsberg, Onsala and Mets\"ahovi. Mainly due
to the high sensitivity of the IRAM telescopes, a detection sensitivity of $\sim (0.2-0.3)$\,Jy
could be reached. This resulted in an impressive detection rate of 95\,\%, which equals to
detection of 121 out of 127 observed sources, of which 109 objects (86\,\%) could be imaged
(see Table \ref{survey}) \citep{Lee07,Lee08}.

%\begin{figure}
%\hspace*{1.0cm} \includegraphics[width=0.50\textwidth,angle=-90]{3C120.epsi}
%\caption{86\,GHz VLBI data of 3C\,120, illustrating the typical data quality and uv-coverage of the new 86\,GHz GMVA
%survey. Data from April 20, 2002 are shown on top, data from Oct. 26, 2002 below. The left panel shows the correlated
%flux density versus projected (uv)-distance. The uv-coverage is plotted as insert. The resulting VLBI maps on the right
%are convolved with an observing beam of (190 x 39)\,$\mu$as (top) and (268 x 36)\,$\mu$as (bottom) and have a
%dynamic range of 67:1 (top) and 32:1 (bottom), respectively \citep{Lee08}.
%}
%\label{3c120a}
%\end{figure}

\begin{figure}[th!]
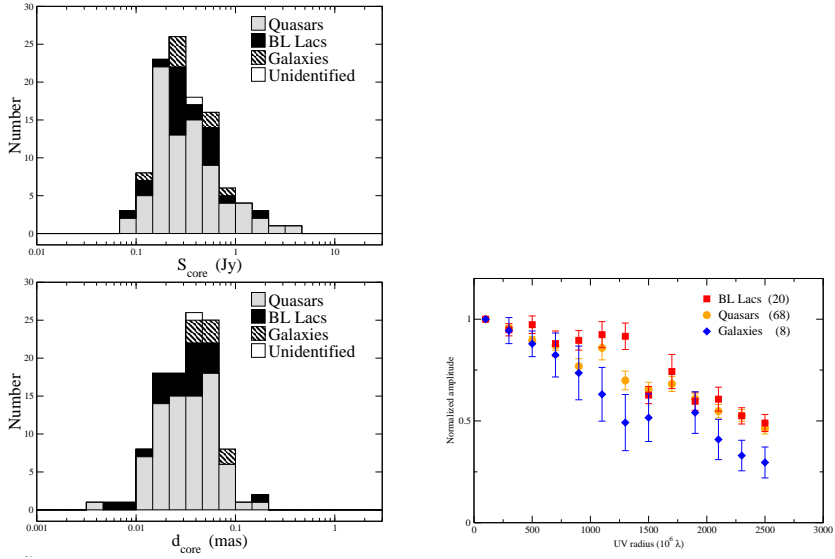

\hspace*{0.5cm} \includegraphics[bb=200 367 429 689,clip=,width=0.40\textwidth, angle=0]{p93.epsi}
\includegraphics[bb=132 364 431 559,clip=,width=0.45\textwidth, angle=0]{p92.epsi}
\caption{Distribution of flux density (top left) and angular size (bottom left) 
of the core components for the survey. The shading denotes different object classes.
Left: Binned, averaged and normalized visibility amplitude plotted versus uv-distance. Different symbols identify 
the different object types. Numbers in brackets give the number of objects in each class. Some
visibility bins (interval: $200$\,M$\lambda$) do not contain data for each source class \citep{Lee08}.
}
\label{histo}
\vspace*{-0.5cm}
\end{figure}

\begin{figure}[h!]
\hspace*{1.0cm} \includegraphics[width=0.30\textwidth,angle=0]{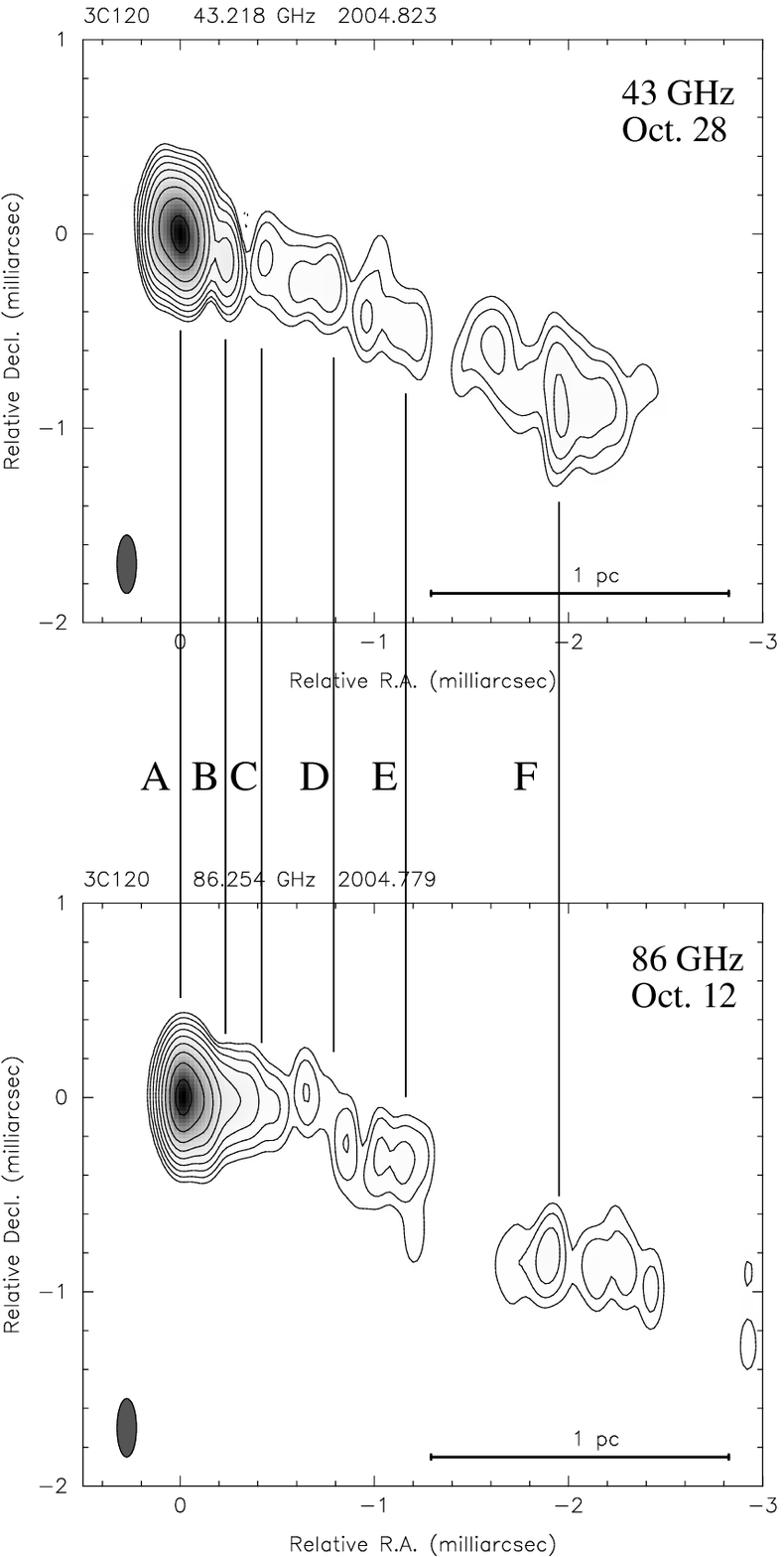}
\hspace*{1.0cm} \includegraphics[width=0.27\textwidth,angle=0]{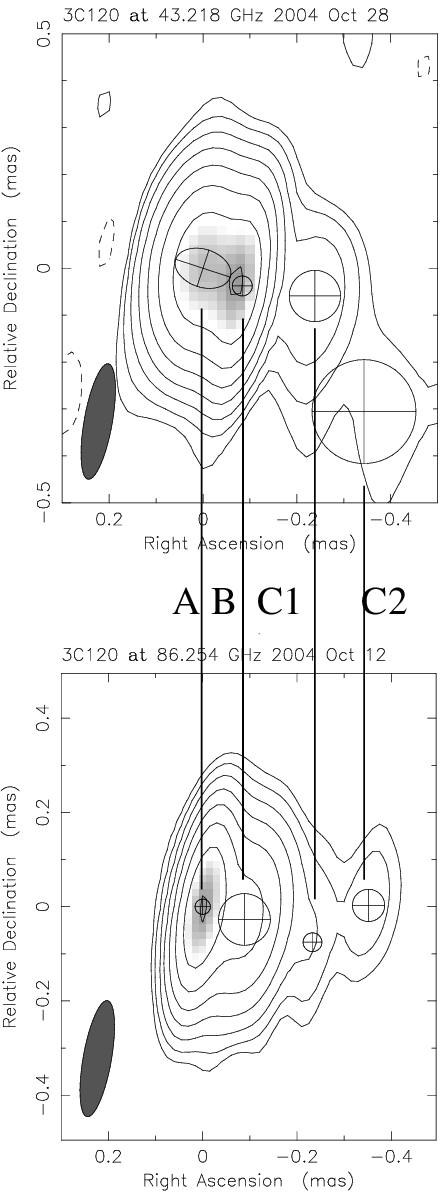}
\caption{VLBI maps of 3C\,120 at 43\,GHz (top, VLBA: Oct. 28, 2004) and 86\,GHz (bottom, GMVA: Oct. 16, 2004).
The maps result from a reanalysis of previously published data \citep{Marscher07}. The
uv-tapered maps on the left show a faint and partially resolved one sided jet extending to at least $r \sim 2$\,mas.
On the right, the central 0.5\,mas region is shown with higher angular resolution (beam: (0.06 x 0.25)\,mas).
Lines and labels guide the eye and help to identify corresponding emission regions. Circles in the contour maps on the 2 right
panels show size and position of Gaussian components fitted to the visibility data. At 86\,GHz, the
core size is $\leq 32\mu$as.
}
\label{3c120b}
\vspace*{-0.5cm}
\end{figure}

%To illustrate some results from this survey, we show in Figure \ref{3c120a} an example of typical visibility data
%and maps. Owing to the snap-shot type observing mode and the resulting limited uv-coverage for each source, 
%the map quality is relatively poor, with typical dynamic ranges of $30-60$. However, in cases where a comparison
%with dedicated full uv-coverage observations was possible, we found a good agreement between the maps from the survey
%and the high dynamic range images from other experiments (e.g compare Fig. \ref{3c120a} with Fig. \ref{3c120b}). 
%In order to parameterize the
%observed brightness distribution, a small number ($N \leq 5$) of Gaussian components was fitted to the visibilities.
%From this the flux, position and size of the individual structural components was estimated. As an example we show in
%Figure \ref{histo} the distribution of the core flux
%densities (top left) and measured VLBI sizes (FWHM) (bottom left) for Quasars, BL-Lacertae Objects, Radio-Galaxies and the 
%optically unidentified objects. Figure \ref{histo} (right panel) shows the mean normalized visibility amplitudes for the optically
%identified sources. 

After fitting of Gaussian components to the visibility data for each 
source, the flux, position and size (FWHM) of the VLBI components were measured.
In Figure \ref{histo} we plot the resulting distribution of the measured core flux
densities (top left) and sizes (bottom left) for Quasars, BL-Lacertae Objects, Radio-Galaxies and the
unidentified objects. The right panel of Figure \ref{histo} shows the mean normalized visibility amplitudes of 
the optically identified sources.

\vspace*{-0.5cm}
\section{The inner jet of 3C\,120}

In Figure \ref{3c120b} we show a 43\,GHz VLBA map and a 86\,GHz GMVA map of the radio galaxy 3C\,120, observed in October 2004.
The left panel shows uv-tapered maps convolved with a (0.1 x 0.3)\,mas beam.
The right panel shows the un-tapered central 0.5\,mas region of the jet, convolved with the smaller 86\,GHz beam 
(0.06 x 0.25\,mas).  At the redshift of 3C\,120 ($z=0.033$) an angular scale of 1\,mas corresponds to a spatial scale
of 0.63\,pc. Assuming negligible variability on the 16 day interval between the two observing dates,
it is possible to determine the spectral indices $\alpha_{\rm 43/86 GHz}$ along the jet ($S \propto \nu^\alpha$).
For the labeled regions A to F in Figure \ref{3c120b} (left), we obtain spectral indices of
-0.35 for A ($r=0$\,mas), -0.07 for B ($r=0.1$\,mas), -0.36 for C ($r=0.25 - 0.4$\,mas), -0.46 for  D ($r=0.8$\,mas),
-0.7 for E ($r=1.2$\,mas) and -0.8 for F ($r=1.8-2.2$\,mas).
A general trend of spectral steepening with increasing core separation is obvious. We note
that component B exhibits a much flatter spectrum than the core component A.  With a size of $\sim 0.1$\,mas (FWHM) 
and a spectral turnover frequency $(43 \leq \nu_m \leq 86)$\,GHz, component B cannot be in energy equipartition,
and is dominated by the magnetic field energy. Similar arguments for the core region (component A, size $\leq 32 \mu$as 
$\leq 7 \cdot 10^{16}$\,cm) and for the outer jet components (C--F) yield no evidence for a substantial deviation from equipartition.
The higher angular resolution at 86\,GHz (Fig. \ref{3c120b}, right) also reveals some structural difference to the 43\,GHz map.
Remarkable is the difference in the position angle of the inner ($r < 0.5$\,mas) jet components. This
could be interpreted as evidence for transverse jet stratification.

\vspace*{-0.5cm}
\section{3C\,454.3 after a major outburst in 2005:}

\begin{figure}[ht!]
\begin{minipage}[t!]{0.47\textwidth}{
\hspace*{0.5cm} \includegraphics[width=0.78\textwidth,angle=0]{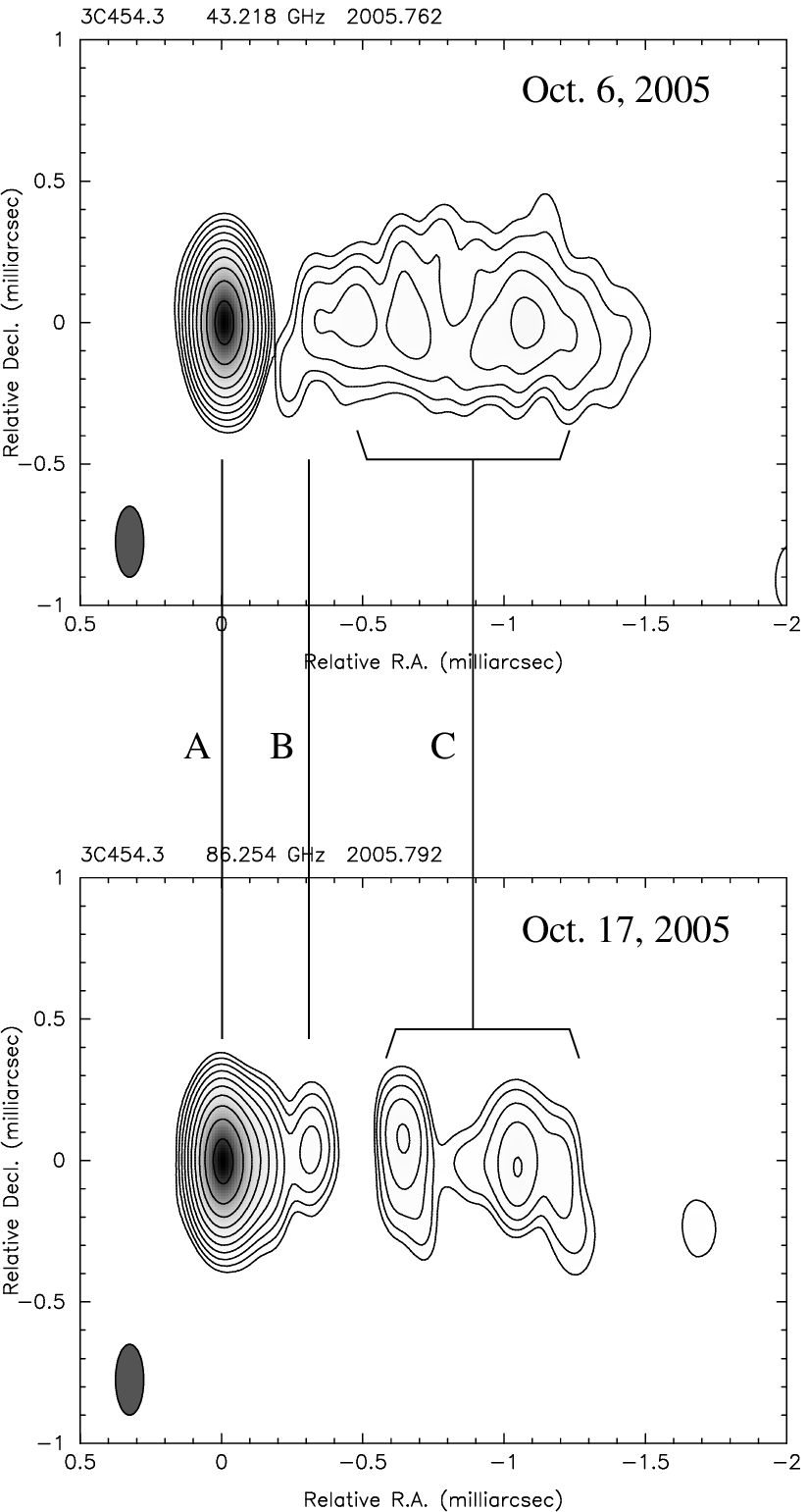}
}
\end{minipage}
~~~
\begin{minipage}[t!]{0.47\textwidth}{
%\vspace*{-0.5cm}
%\includegraphics[width=0.75\textwidth,bb=64 33 576 720,clip,angle=-90,origin=br]{sma_3c454.3.epsi}
\includegraphics[width=0.63\textwidth,bb=64 33 576 720,clip,angle=-90,origin=br]{sma_3c454.3.epsi}
~~\\
~~\\
~~\\
~~\\
\hspace*{-0.5cm}\includegraphics[width=0.94\textwidth,angle=0]{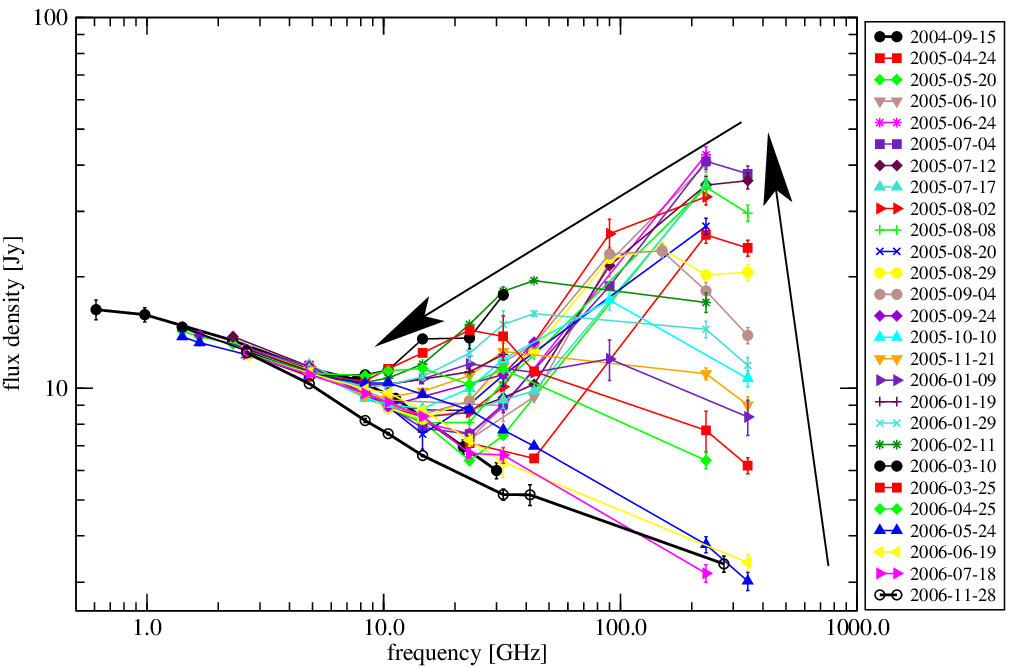}
}
\end{minipage}
\caption{Left: 43\,GHz VLBA (top, data from Marscher et al.) and 86\,GHz GMVA (bottom) map of 3C\,454.3 observed in Oct.\ 6 
and Oct.\ 17, 2005.  Both maps are restored with a common beam of (0.1 x 0.25)\,mas.
Lines with labels connect the three corresponding emission regions A, B and C (see text).
Top right:  Flux density variability at 230 and 345\,GHz. The insert shows the variability during the time of the peak 
flux  \citep[data: SMA,][]{Gurwell07}.
Bottom right: Evolution of the radio spectrum of 3C\,454.3 after the optical flare. The spectra cover a time range
from Sep.\ 2004 until Nov.\ 2006 at time intervals of a few weeks. The arrows mark the direction of
spectral evolution. Each spectrum is obtained from quasi-simultaneous flux measurements, combining data 
from Effelsberg, VLA, IRAM, SMA \& Ratan-600.
}
\label{3c454.3}
\end{figure}

\begin{figure}[t!]
\hspace*{0.5cm}\includegraphics[angle=-90,width=0.9\textwidth]{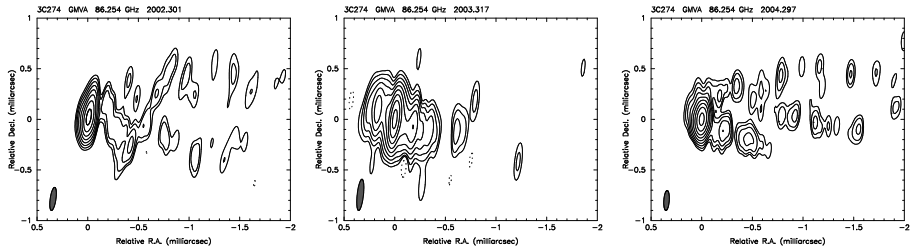}
\caption{86\,GHz VLBI images of M\,87 (3C\,274) obtained with global 3\,mm VLBI
in April 2002, 2003, and 2004 (from left to right).
The contour levels are at -0.3, 0.3, 0.6, 1.2, 2.4, 4.8, 9.6, 19.2, 38.4, and 76.8 \% of the peak flux, of
0.52 Jy/beam (2002), 0.73 Jy/beam (2003), 0.35 Jy/beam (2004). The lowest contour
level is omitted in the image of 2002. The restoring beam sizes are $0.23 \times 0.057$\,mas (2002), $0.31 \times 0.062$\,mas (2003),
and $0.20 \times 0.054$\,mas (2004), respectively.
}
\label{m87}
\vspace{-0.5cm}
\end{figure}

In May-June 2005, the quasar 3C\,454.3 ($z=0.859$) showed a major flux density outburst, which was observed 
from hard X-ray to radio bands 
%\citep[e.g.][and references therein]{Villata07}. 
(e.g. Villata et al.\ 2007, and references therein).
The flux variations
at 1.3\,mm and 0.85\,mm are shown in Figure \ref{3c454.3} (top, right). During the flare maximum
quasi-periodic oscillations on a $\sim 11$\,day time scale appear, possibly indicating motion in a helical jet
with $\sim 100$ Schwarzschild radii diameter (assuming $M_{\rm BH} =10^9 M_\odot$). 
A Doppler-boosting factor of $\delta \simeq 5-9$ would be necessary to bring the observed 
variability brightness temperature of $T_B \simeq 10^{14}$\,K (at 230\,GHz) 
back to the inverse Compton limit of $\sim 10^{12}$\,K.  The spectral
evolution of the flare through the radio bands is shown in Figure \ref{3c454.3} (bottom, left). It is obvious that
the variability is most pronounced at short mm-wavelengths, with a spectral turnover moving from initially  $\simeq  230$\,GHz
to lower frequencies. Below $\sim 10$\,GHz almost no variability is seen.

So far, VLBI observations at frequencies $\nu \leq 43$\,GHz did not reveal strong evidence for the ejection of
a new jet component, which could be related to this flare. In Figure \ref{3c454.3} (left panel) we show two VLBI maps obtained
about 4 months after the flare. The 86\,GHz data reveal a clear elongation of the core region A, which is
not visible in the 43\,GHz map (region B). A spectral index calculation for the three main
emission regions yields 
for the core A $\alpha_{\rm 43/86 GHz} = -0.4$, for the region B $\alpha = +(1.1 ... 2.6)$
and for the jet C $\alpha = -0.3$. The strongly inverted spectrum of the region B indicates possible
fore-ground absorption, perhaps from a cloud or torus. With a radial extend ranging from 
$r_{\rm min} \la 0.1$\,mas to $r_{\rm max} \simeq 0.3$\,mas (0.1 mas corresponds to 0.75\,pc) 
and a sufficiently high opacity, the inner portion of the jet would become visible only at the highest frequencies.\\

%\vspace*{-0.8cm}
%\section{The Jet of M\,87}
\noindent
{\bf 5.~~ The Jet of M\,87}\\
~~\\
In Figure \ref{m87} we show 3 images of the inner jet in M\,87 (3C\,274) obtained with the GMVA at 86\,GHz during 2002-2004.
The map of 2004 has a higher dynamic range than the earlier images due to increased observing bandwidth (512\,Mbps).
To our knowledge, these are the highest angular and spatial resolution maps ever made of this source. At a distance of
D=16.75\,Mpc, the minor axis of the GMVA observing beam of $\sim 54 \mu$as translates into a spatial scale
of $1.35 \cdot 10^{16}$\,cm (4.4\,mpc). From the slightly elliptical observing beam one obtains a formal upper limit for the core 
size of $\leq$(200 \rm x 54)\,$\mu$as, which corresponds to $R_s$=(56 x 15) Schwarzschild radii (for
$M_{BH}=3 \cdot 10^9 M_\odot$). We may identify the size of the brightest and most compact VLBI component 
with the jet diameter at or near its origin.
%In this case it is remarkable that the jet base is so small and bright ($T_B \geq 2 \cdot 10^{10}$\,K), 
%contrary to expectations from magnetic-slingshot type acceleration models \citep[e.g.][]{BlandfordPayne82} with magnetic
%fields anchored in the rotating accretion disk and a jet width larger than twice
%the light-cylinder radius \citep[$\geq \gamma\cdot 2 R_L$,][]{Camenzind90}.
%The observed small size therefore points more towards models
%in which the jet is attached more directly to the rotating black hole \citep{BlandfordZnajek77}, gaining its
%energy via MHD interaction with the inner disk \citep[e.g.][]{McKinney06}
%or the Penrose process \citep[][and references therein]{Gariel07}.
In this case it is remarkable that the jet base is so small and bright ($T_B \geq 2 \cdot 10^{10}$\,K), 
contrary to expectations from magnetic-slingshot type acceleration models \citep[e.g.][]{BlandfordPayne82}. Here,
magnetic fields are anchored in the rotating accretion disk, expand and form a light-cylinder. Its
diameter defines the minimum jet width of $> 50 R_s$ \citep{Camenzind90,FendtMemola01}.
%diameter defines the minimum jet width of $> 50 R_s$ \citep{FendtMemola01}.
The observed small size therefore points more towards models
in which the jet is attached more directly to the rotating black hole \citep{BlandfordZnajek77}, gaining its
energy via MHD interaction with the inner disk \citep[e.g.][]{McKinney06}
or the Penrose process \citep[][and references therein]{Gariel07}.

The images of 2002 and 2004 both show a bifurcated conically opening
one-sided jet, similar to the jet morphology seen at 43\,GHz (Walker et al., this conference). 
The transverse width of the jet is of order of $0.5-0.7$\,mas, corresponding to a jet diameter of $\sim 140-200$
Schwarzschild radii (at $r=0.5-2$\,mas core separation). There are clear signs of edge-brightening 
and a `hollow' or at least faint central jet body. We note that
for a jet inclination $> (30-45)$\deg, a fast jet spine could be Doppler-deboosted and remain faint.

In 2003, a component located $\sim 0.2$\,mas east of a $\sim 4$ times brighter and more compact component is seen.  
At present it is unclear, if this eastern component should be identified with the jet base (which then must 
vary in compactness), or if the eastern component is part of a counter-jet. In any case, the structural differences 
between the 3 maps (and in the corresponding visibility data) indicate pronounced jet variability on and below the
1 year timescale. For the variations with an angular rate $\geq 0.2$\,mas/yr the corresponding jet speed would be 
$\beta_{\rm app} > 0.05$. More frequent global 3mm-VLBI observations are required to follow this motion.\\

%\vspace*{-0.5cm}
%\section{VLBI at shorter wavelengths and Future Outlook}
\noindent
{\bf 6.~~ VLBI at shorter wavelengths and Future Outlook}\\
~~\\
After the successful detection of a number of radio source on continental and on transatlantic VLBI baselines 
at 129, 147 and 230\,GHz \citep[see][and references therein]{Krichbaum07}, the technical feasibility of
VLBI at wavelengths shorter than 3\,mm has now been demonstrated. In the next step, and after upgrade of the existing
VLBI data acquisition systems to higher data rates of $\sim 2-4$\,Gbps and observing bandwidths of $\geq 512$\,MHz, the 
baseline sensitivity will dramatically improve ($\la 0.5$\,Jy at 230\,GHz for the existing antennas). 
With a relatively small VLBI array consisting of $N=2-4$ mm-telescopes,
a systematic VLBI survey of the degree of compactness of a large sample of radio sources could be tackled. 
%The further addition of more and more sensitive mm-capable telescopes, 
The further addition of more and more sensitive mm-capable telescopes (see Table \ref{antenna}), 
will lead to true imaging capability\footnote{At least 4-5 stations are needed for hybrid mapping; the image quality 
increases with the number of VLBI telescopes.}. 
Such effort appears justified, since the VLBI imaging of the Galactic Center source Sgr\,A*, M\,87 and other 
nearby objects at 230 and 345\,GHz will allow to map the direct vicinity of super massive black holes 
with a spatial resolution of $\leq 5-6$ Schwarzschild radii !
In this context, the participation of the large and sensitive millimeter interferometers (SMA, PdB, CARMA, ALMA) and 
large single dish telescopes (e.g the LMT) is essential, not only for reaching a high detection sensitivity, but also
for providing a dense uv-coverage. Only with the latter an image fidelity can be reached,
which in the end might allow to directly observe the expected image distortions and General Relativity effects in 
the vicinity of a black hole.

\noindent
\begin{table}
{\small 
\begin{tabular}{lllllcccc}
Name             & Country  & D       & Alt.     & Surf.     &  90         &    150       &  230       &   345      \\
                 &          & [m]     & [m]      & [$\mu$m]  &  [GHz]      &    [GHz]     &  [GHz]     &   [GHz]    \\ \hline       
                 &          &         &          &           &             &              &            &            \\ 
\multicolumn{2}{l}{Surface $< 50\mu$m}&          &           &             &              &            &            \\     
SMA              & HI, USA  & 8x6     &4100      &  12       &     -       &      -       &   +        &    +       \\    
HHT-SMTO         & AZ, USA  & 10      &3100      &  15       &     -       &      V       &   V        &    +       \\                 
APEX             & Chile    & 12      &5000      &  18       &     -       &      -       &   +        &    +       \\      
ASTE             & Chile    & 10      &4800      &  20       &     -       &      ?       &   ?        &    +       \\     
CSO              & HI, USA  & 10      &4100      &  25       &     -       &      ?       &   +        &    +       \\     
JCMT             & HI, USA  & 15      &4100      &  25       &     -       &      ?       &   V        &    +       \\     
                 &          &         &          &           &             &              &            &            \\ 
\multicolumn{2}{l}{Surface $50-100\mu$m}&        &           &             &              &            &            \\  
Pl.de Bure       & France   & 6x15    &2550      &  55       &     V       &      +       &   V        &    +       \\                    
Pico Veleta      & Spain    & 30      &2900      &  67       &     V       &      V       &   V        &    +       \\                           
CARMA            & CA, USA  & $^a$    &2200      & 30-60     &     +       &      ?       &   V        &    +       \\       
Kitt Peak        & AZ, USA  & 12      &2000      &  75       &     V       &      V       &   +        &    -       \\            
                 &          &         &          &           &             &              &            &            \\ 
\multicolumn{2}{l}{Surface $100-300\mu$m}&       &           &             &              &            &            \\  
Metsahovi        & Finland  & 14      &sea l.    &  100      &      V      &       V      &    -       &     -      \\                    
Mopra            & Australia& 22      &900       & $\sim$120 &     +       &      -       &   -        &    -       \\      
Onsala           & Sweden   & 20      &sea l.    &  130      &      V      &       ?      &    -       &     -      \\             
VLBA             & USA      & 25      &$^b$      &  150      &      V      &       ?      &    -       &     -      \\
Taeduk           & Korea    & 14      & 200      & 170       &     +       &      ?       &   -        &    -       \\     
Nobeyama         & Japan    & 45      &1300      & 200       &     +       &      +       &   -        &    -       \\     
Cambridge        & UK       & 32      &  20      & 200       &     ?       &      -       &   -        &    -       \\     
Haystack         & MA, USA  & 37      & 150      & 250       &     V       &      -       &   -        &    -       \\                 
Delingha         & China    & 14      &3200      & 250       &     +       &      -       &   -        &    -       \\        
Simeiz           & Ukraine  & 22      &sea l.    & 250       &     ?       &      -       &   -        &    -       \\     
                 &          &         &          &           &             &              &            &            \\ 
\multicolumn{2}{l}{Surface $300-500\mu$m}&       &           &             &              &            &            \\  
GBT              & VA, USA  & 100     &840       & 390$^c$   &     +       &      -       &   -        &    -       \\           
Noto             & Italy    & 32      & 30       & 400       &     +       &      -       &   -        &    -       \\     
Effelsberg       & Germany  & 100     &300       & 450       &     V       &      -       &   -        &    -       \\            
                 &          &         &          &           &             &              &            &            \\ 
\multicolumn{3}{l}{under construction}           &           &             &              &            &            \\ 
ALMA             & Chile    & 50x12   &5000      &  25       &     +       &      +       &   +        &    +       \\     
LMT              & Mexico   & 50      &4600      &  70       &     +       &      +       &   +        &    +       \\     
Haystack$^d$     & MA, USA  & 37      & 150      &  100      &     +       &      ?       &   -        &    -       \\ 
Yebes            & Spain    & 40      & 850      & 150       &     +       &      ?       &   -        &    -       \\     
KVN              & Korea    & 3x21    & 200      &$\sim$150  &     +       &      ?       &   -        &    -       \\      
SRT              & Italy    & 64      & 600      &$\sim$200  &     +       &      ?       &   -        &    -       \\      
\end{tabular}
\begin{tabular}{ll}
      & \\
\multicolumn{2}{l}{Notes:} \\
$^a$:&inhomogeneous array (6x10m+9x6m)                                            \\
$^b$:&10 different antennas with elevations ranging from sea level to 4100\,m    \\
$^c$:&present value, aim is $\sim 210\mu$m                                       \\
$^d$:&after upgrade                                                              \\
\end{tabular}
}

\caption{
Antennas for mm-VLBI, present status and future candidates. Col. 3 gives the antenna diameter, col. 4 the
altitude, col. 5 the surface rms. Columns 6-9 summarize the suitability for mm-VLBI at a given observing 
band, with the following symbols: 
"V" : VLBI successfully done, receiver available;
"+" : in principle possible, receiver available or planned;
"?" : perhaps possible, but presently no receiver;
"-" : not possible, surface not good enough.
}
\label{antenna}
\end{table}

%\acknowledgements %%% Text of acknowledgments runs on after this command.

%%% THE BIBLIOGRAPHY
%%%
%%% CONSULT SECTION 3 OF "INSTRUCTIONS FOR AUTHORS" FOR HOW TO USE NATBIB.
%%% AUTHORS ARE ENCOURAGED TO USE EITHER THE "THEBIBLIOGRAPY" ENVIRONMENT
%%% BY UNCOMMENTING (DELETING THE "%" SYMBOL) THE COMMANDS BELOW, OR BY
%%% USING THE BIBTEX ENVIRONMENT. TO FIND OUT WHICH IS APPLICABLE TO YOUR
%%% CONTRIBUTION, CONSULT THE VOLUME EDITORS FOR YOUR PROCEEDINGS.
%%%

\end{document}